# Low-noise 750-MHz spaced Yb:fiber frequency combs

YUXUAN MA,[1,2,†] BO XU,[1,3,†] HIROTAKA ISHII,[1] FEI MENG,[2] YOSHIAKI NAKAJIMA,[1,3] ISAO MATSUSHIMA,[1,3] THOMAS R. SCHIBLI,[4] ZHIGANG ZHANG,[2] AND KAORU MINOSHIMA[1,3,*]

[1]Department of Engineering Science, Graduate School of Informatics, The University of Electro-Communications (UEC), 1-5-1 Chofugaoka, Chofu, Tokyo 182-8585, Japan
[2]State Key Laboratory of Advanced Optical Communication System and Networks, School of Electronics Engineering and Computer Science, Peking University, Beijing 100871, China
[3]Japan Science and Technology Agency, ERATO MINOSHIMA Intelligent Optical Synthesizer (IOS) Project, 1-5-1 Chofugaoka, Chofu, Tokyo 182-8585, Japan
[4]Department of Physics, University of Colorado at Boulder, 2000 Colorado Avenue, Boulder, Colorado 80309-0390, USA
*Corresponding author: k.minoshima@uec.ac.jp

**Growing interest of frequency combs is stimulating the development of high repetition rate combs. However, increasing the repetition rate to ~GHz levels while maintaining tight phase locking to an optical reference with <1 rad RMS phase is still challenging for fiber-based frequency combs. In this paper, we demonstrate two low-noise 750-MHz ytterbium (Yb) fiber frequency combs that are independently stabilized to a continuous wave (CW) laser. A bulk electro-optic modulator (EOM) and a single-stack piezo-electric transducer (PZT) are employed as fast actuators for stabilizing the respected cavity length to heterodyne beat notes ($f_{beat}$). Both combs exhibit in-loop fractional frequency instabilities of ~$10^{-18}$ at 1 s. This is the first demonstration of tightly phase-locked fiber frequency combs with 750 MHz fundamental repetition rate. The experiment shows that either the thin EOM or single stack PZT is a practical scheme for developing low-noise high-repetition-rate Yb:fiber combs. Those combs will be powerful tools for advanced applications such as optical frequency atomic clocks and pulse stacking.**

**OCIS codes:** (320.7090) Ultrafast Lasers; (140.3510) Lasers, fiber; (120.3930) Metrological instrumentation.

The development of optical frequency combs (OFCs) over the past decade has experienced fascinating evolution toward high frequency stability, high coherence, high repetition rate, and high output power. This is particularly evident when we look at the wide variety of OFCs-based numerous exciting applications such as ultra-low noise microwave generation [1, 2], optical waveform synthesizers [3, 4], astronomical calibration [5], and dual-comb spectroscopy [6]. Elevating the repetition rate to ~GHz levels on the premise of low noise will have further benefits on increasing the scan speed used in dual-comb distance measurements [7], improving the sensitivity of dual-comb spectroscopy [8], and extending the measurable range of one-shot 3D shape imaging [9]. Multiple configurations of frequency combs with ~GHz repetition rates, including solid-state lasers, fiber lasers, and micro-resonators have been reported [10-15]. High-repetition-rate Yb frequency combs are prominent owing to their exceptional characteristics such as superior power scalability and feasible Si-photo detector measurements.

Phase locking the comb to an optical frequency reference other than a radio frequency (RF) standard can reduce the phase error to < 1 rad and narrow the linewidth to sub-mHz levels [16]. This is propitious to increase frequency stability by four to six orders of magnitude in, for example, an 800 MHz Ti:sapphire laser [17]. However, similar phase-locking records in fiber-laser-based frequency combs have only been obtained at a relatively low repetition rate (typically around 100 MHz) [16, 18, 19].

The tight phase locking in high-repetition-rate Yb:fiber frequency combs involves two issues. One is the insufficient servo bandwidth against high intrinsic noise, and the other is crosstalk between the intracavity electro-optic modulator (EOM) used to stabilize the cavity length and gain modulation to $f_{ceo}$.

Larger servo bandwidths are required because the intrinsic noise in $f_{ceo}$ generally scales up with the repetition rate [10, 11]. Therefore, it is necessary to achieve ~MHz-level servo bandwidth for $f_{ceo}$ with >100 kHz free running linewidth, which is a common characteristic of Yb:fiber lasers with high repetition rate. However, the gain modulation bandwidth is limited by the excited-state life time of the gain material and the available pump laser diode (LD) driver. A promising solution is to employ loss modulation instead of gain modulation. Graphene modulators are capable of producing large modulation bandwidths and have been used to successfully lock $f_{ceo}$ with ~1 MHz servo bandwidth in a Thulium fiber comb [20] and with 1.6 MHz servo bandwidth in an Erbium fiber comb [19] respectively. However, if the graphene modulator is applied to the ~1 micron wavelength range, the doping concentration required will be too high to reach the desired transparency. On the other hand, it was also reported that gain modulation bandwidth of more than a few hundred kHz has been achieved in an Er:fiber comb by directly feeding back the modulated current to the pins of the pump LD [18]. However, only a few publications in the literature report Yb:fiber combs with tight phase locking of $f_{ceo}$ [16, 21]. Those reported combs benefit from the record free-running narrow linewidth (<10 kHz) of $f_{ceo}$ for fiber lasers.

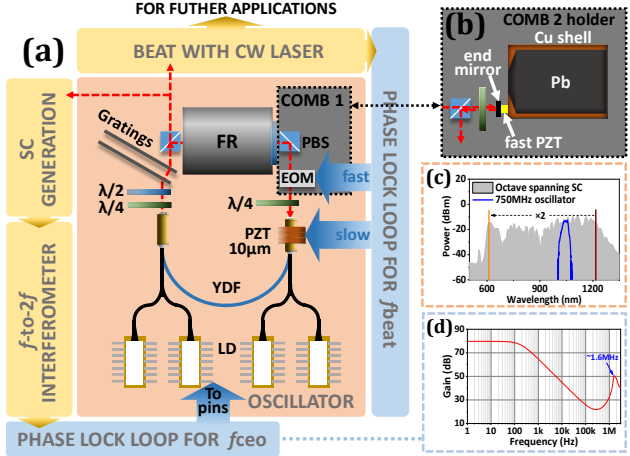

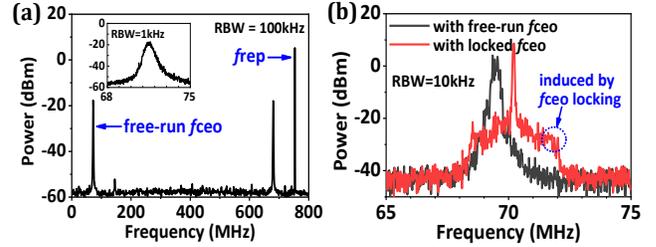

**Fig. 1.** (a) Configuration of Comb 1. (PBS: polarization beam splitter, YDF: ytterbium doped fiber). (b) Structure of the fast PZT section in Comb 2. (c) Octave-spanning SC generation (blue curve: optical spectrum of the 750 MHz oscillator). (d) Open-loop response of the PID circuit for locking $f_{ceo}$.

**Fig. 2.** (a) RF spectrum of free-running $f_{ceo}$ signal. Inset shows the zoom-in view; (b) Free-running $f_{beat}$ signals with free-running (grey curve) and locked (red curve) $f_{ceo}$.

Cross-influence occurs when an intra-cavity EOM is involved. Minute misalignment of the EOM crystal causes coupling between cavity length modulation and pulse polarization via the complex laser dynamics [22], hence introducing competition between the stabilizations of $f_{ceo}$ and $f_{beat}$. Dynamic mounting is supposed to be able to solve the problem, but resonance ringing is not avoidable. An alternative candidate to overcome this drawback is to use a single-stack piezo actuator that can potentially operate with a response bandwidth of a few hundred kHz [23].

We presented a fully stabilized 750-MHz Yb:fiber comb phase locked to both microwave standards and optical references in a previous paper [24]. However, the residual phase errors of $f_{ceo}$ and $f_{beat}$ were still high (13.7 rad and 4.6 rad, respectively), and the in-loop instability was $10^{-15}$ at 1 s. In this work, we demonstrate phase-stabilized 750-MHz Yb:fiber frequency combs with <1 rad RMS phase both in $f_{ceo}$ and $f_{beat}$ in two distinct configurations. Both combs exhibit in-loop fractional frequency instability of $10^{-18}$ at 1 s. To the best of our knowledge, this is the first report of tightly phase-locked fiber-based combs with the highest fundamental repetition rate of 750 MHz.

We denote Comb 1 as a Yb:fiber comb equipped with an intra-cavity EOM (a 2-mm-long, z-cut LiNbO$_3$ crystal) in transmission configuration, and the other comb with a single-stack PZT (PL022.3x, Physik Instrumente GmbH) in reflection configuration as Comb 2. The slow PZTs with 10 μm travel range are separately responsible for drift below 2 kHz in both the combs.

Comb 1 is based on the ring-cavity 750-MHz Yb:fiber laser described in [24], as shown in Fig. 1(a). The major output power (~630 mW) of the oscillator was directly launched into a home-made tapered photonic crystal fiber (PCF) for generating an octave spanning super-continuum (SC) spectrum (Fig. 1(c)). A free-running $f_{ceo}$ signal (RBW=100 kHz, plotted in Fig. 2(a)) with >40 dB signal-to-noise ratio (SNR) was obtained using a conventional f-to-2f interferometer. A small portion (~5 mW) of the output power was used to beat with the 1064 nm CW laser (a monolithic isolated single-mode end-pumped ring laser with approximately 3 kHz free-running linewidth). In the servo loops, the generated $f_{ceo}$ and $f_{beat}$ signals were separately filtered, electronically amplified, and directly phase locked to the 70 MHz frequency references using digital phase discriminators and home-made PID loop filters. Several considerations are given in order to realize phase locking with <1 rad RMS phase in the 750 MHz Yb:fiber combs.

First, an optimum mode-locking status in the stretched pulse regime with an appropriate LD bias current is an indispensable precondition. Only in this way can it be possible to acquire a maximized linear $f_{ceo}$ tuning rate ($\Delta f_{ceo}/\Delta I_{LD}$=2.1 MHz/mA in this experiment) while simultaneously maintaining invariable free-running linewidth (~200 kHz) and SNR within a large current tuning range (~800 mA). This is a key factor since we found it was difficult to observe the coherent peak locked to $f_{ceo}$ if the oscillator operated at a terribly low $f_{ceo}$ tuning rate (e.g., <0.3 MHz/mA).

Second, employing digital phase discriminators instead of frequency dividers favors authentically revealing the unaltered features of the beat signals. This can prevent degradation of frequency instability induced by frequency division.

Third, two home-made, high-speed PID loop filters must be specifically tailored to the high-repetition-rate Yb:fiber frequency combs. For the $f_{ceo}$ loop filter, two differentiator stages were employed to compensate the descending current modulation response at high frequency. The increasingly high open loop gain was cut-off at ~1.6 MHz (as shown in Fig. 1(d)), which is limited by the gain-bandwidth product (GBP) of the specific operational amplifiers. The aforementioned single servo loop is capable of tightly phase locking $f_{ceo}$. With respect to $f_{beat}$, the voltage applied to the EOM ranged within 0 V ±1 V, thus cooperating with the slow PZT for cavity length stabilization throughout the entire locking period. The outputs from the PID circuits directly feed back to the LD electrodes and the EOM electrodes to maintain the desired bandwidth.

The RF spectra and phase noise spectra of simultaneously phase locked $f_{ceo}$ and $f_{beat}$ of Comb 1 are shown in Fig. 3. The insets in Figs. 3(a) and 2(c) show the RBW-limited linewidth for both signals, indicating coherent peaks with 50 dB (RBW = 300 Hz) SNR, which are equivalent to ~75 dB×Hz. It is noted that the stabilized $f_{ceo}$ shows a servo bump at 1.6 MHz, which is recognized as a consequence of the GBP in the PID circuits explained previously. Obvious dents at ~900 kHz in both Figs. 3(a) and 3(c) are induced by the strong EOM resonance applied to both $f_{ceo}$ and $f_{beat}$. Figures 3(b) and 3(d) show considerable phase noise reductions in both signals after stabilization. The residual RMS phase errors are respectively calculated to be 0.49 rad for $f_{ceo}$ and 0.35 rad for $f_{beat}$ with 0.1 Hz to 10 MHz integration range. This indicates energy concentrations of 81% and 89% in the corresponding coherent peaks, respectively. Both $f_{ceo}$ and $f_{beat}$ exhibit in-loop fractional frequency instability of ~1.5×10$^{-18}$ with an averaging time of 1 s and an improved frequency instability of ~2.5×10$^{-20}$ at 100 s averaging time (counted by Agilent 53132A and normalized to 282 THz, as shown in Fig. 3(e)). All the RF frequency synthesizers, frequency counters, and phase noise analyzers are referenced to a GPS-disciplined Rubidium (Rb) oscillator. One should note that there is a distinct rising tendency when the EOM was enabled (see Fig. 3(b)). This is caused by crosstalk between the EOM and $f_{ceo}$. Since crosstalk can severely deteriorate the locked $f_{ceo}$ if the EOM bandwidth is fully utilized, a trade-off must be made by keeping the cut-off frequency of $f_{beat}$ servo loop at ~120 kHz, although the resonance frequency of EOM is ~900 kHz.

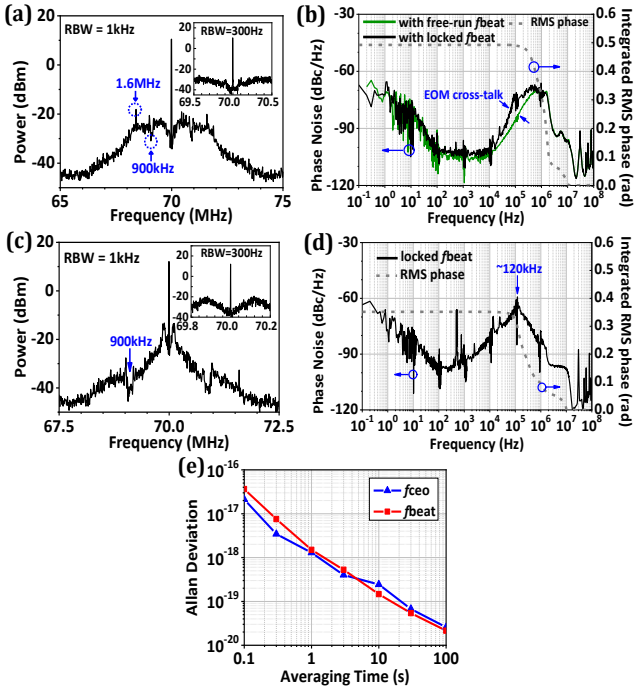

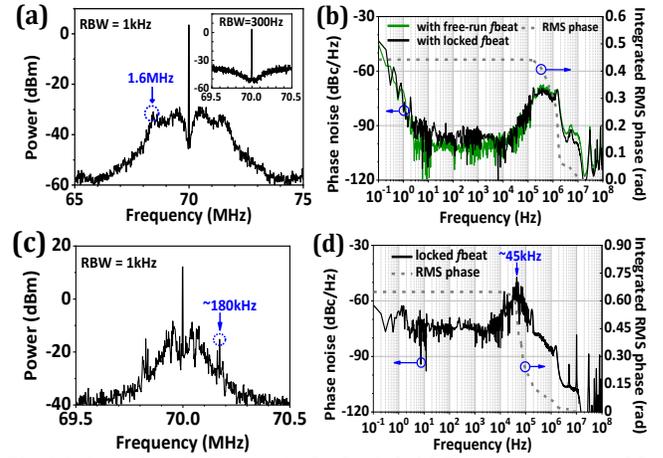

**Fig. 3.** In-loop phase locking results for Comb 1: (a) In-loop $f_{ceo}$ with simultaneously stabilized $f_{beat}$. The inset shows a magnified view with smaller span and RBW. (b) Phase noise spectra of $f_{ceo}$ (phase locked $f_{ceo}$ with free-running $f_{beat}$: green solid; and simultaneously stabilized $f_{beat}$: black solid) and integrated phase error of $f_{ceo}$ under full phase locking (gray dash). (c) In-loop $f_{beat}$ with simultaneously stabilized $f_{ceo}$. The inset shows a magnified view. (d) Phase noise spectra of $f_{beat}$ (full stabilization: black solid) and the integrated phase error of $f_{beat}$ (grey dashed curve). (e) Relative Allan deviations (normalized to 282 THz) of the in-loop $f_{ceo}$ (blue triangle) and $f_{beat}$ (red square) under full stabilization.

In order to reduce crosstalk originating from the intracavity components, we built Comb 2, which employs a high-speed single-stack PZT (fast PZT) for stabilizing the cavity length. Accordingly, the cavity must be modified as "σ" type so that the laser can have a linear path for the implementation of the end mirror driven by the fast PZT. The configuration of Comb 2 is schematically illustrated by simply replacing the grey highlighted area in Fig. 1(a) with Fig. 1(b) and leaving the remaining structures unchanged. It is believed that such a modification scarcely influences the optical characteristics of the oscillator; thus, both the free-running $f_{ceo}$ and $f_{beat}$ signals have nearly the same behavior as in Comb 1.

To exploit the highest response bandwidth of the fast PZT, we engineered a special holder following the technique introduced in Ref. [23]. As shown in Fig. 1(b), the PZT holder is a hollow copper (Cu) cylinder filled with lead (Pb) to dampen longitudinal resonance waves. One end inside the cylinder was cone-shaped to prevent acoustic resonance. At the same time, the surface of the fast PZT was roughened with a diamond saw for the same reason. A 3×3×1 mm³ tiny end mirror and fast PZT were bonded to the edge of the holder with a rigid adhesive to eliminate the drum effect. All these procedures are required to push the lowest resonance frequency of the PZT to ~180 kHz. The PID parameters of the $f_{beat}$ servo loop were adjusted according to the response of the fast PZT, and the applied voltage was maintained within 0 V ±10 V. The fast PZT could maintain hour-level locking without the help of a slow loop.

The phase-locking results of Comb 2 are shown in Fig. 4. Of great importance, the stabilized $f_{ceo}$ spectrum was no longer affected by the $f_{beat}$ locking loop (Fig. 4(b)); therefore, $f_{ceo}$ behaves a smoother RF spectrum and has a lower integrated residual phase error than Comb 1 (Fig. 4(a). Two peaks at ~180 kHz were observed, as shown in Fig. 4(c). These peaks are identified as the resonance of the fast PZT. To avoid this, the effective servo bandwidth was bound to ~45 kHz (see Fig. 4(d)). The calculated RMS phase errors are 0.44 rad for $f_{ceo}$ and 0.65 rad for $f_{beat}$ (corresponding to 84% and 70% energy concentrations), respectively. Both in-loop frequency Allan deviations of Comb 2 are at the same level (not shown) as Comb 1.

In order to evaluate the time-domain properties of the frequency combs, we deduced timing jitters from the measured phase noise spectra according to the formula given in [25],

$$S_{\Delta\nu_N \Delta\nu_N} = S_{\Delta f_{ceo}\Delta f_{ceo}} + S_{N\cdot\Delta f_{rep} N\cdot\Delta f_{rep}} + \Gamma_\Delta(\omega) \times \sqrt{S_{\Delta f_{ceo}\Delta f_{ceo}} \cdot S_{N\cdot\Delta f_{rep} N\cdot\Delta f_{rep}}} \quad (1)$$

where $S_{\Delta\nu_N \Delta\nu_N}$, $S_{\Delta f_{ceo}\Delta f_{ceo}}$, and $S_{N\cdot\Delta f_{rep} N\cdot\Delta f_{rep}}$ are the frequency noise power spectral densities (PSDs) of $f_{beat}$, $f_{ceo}$, and $N\cdot f_{rep}$; $\Gamma_\Delta(\omega)$ is the sum of coherence. Since we observed a much narrower free-running $f_{beat}$ (Fig. 2(b), which will be discussed in the following paragraph) after phase locking $f_{ceo}$, which is similar to the anti-correlated condition in Ref. [25], here we assume $\Gamma_\Delta(\omega) = -2$ for simplicity. N is the mode number, which was determined to be ~376,000 in the 750 MHz frequency comb. Dividing the calculated $S_{N \cdot \Delta f_{rep} N \cdot \Delta f_{rep}}$ by $N^2$, we can obtain the PSD jitter of $f_{rep}$ and the corresponding integrated timing jitters (Fig. 5). We find that jitters below 1 kHz are heavily suppressed to lower than 50 atto-seconds in both combs, and the total jitters integrated from 0.1 Hz to 10 MHz are 4.9 fs for Comb 1 and 5.3 fs for Comb 2. The integration frequency range of the noise spectra contributes differently to the jitters in each comb. The jitter in Comb 1 is primarily contributed by $10^5$–$10^6$ Hz noise, where the competition between $f_{ceo}$ and $f_{beat}$ raises the noise level even higher than that in the free-running signal. In contrast, the jitter of Comb 2 is primarily contributed by tens of kHz noise due to the narrower response bandwidth of the PZT than that of the EOM.

**Fig. 4.** In-loop phase locking results for Comb 2: (a) Phase noise spectra of $f_{ceo}$. Inset shows a magnified view. (b) Phase noise spectra of $f_{ceo}$ (only locked $f_{ceo}$: green solid; full phase locking: black solid) and integrated phase error of $f_{ceo}$ (gray dash). (c) RF spectrum of in-loop $f_{beat}$. (d) Phase noise spectra of $f_{beat}$ (full phase locking: black solid) and the integrated phase error of $f_{beat}$ (grey dashed curve).

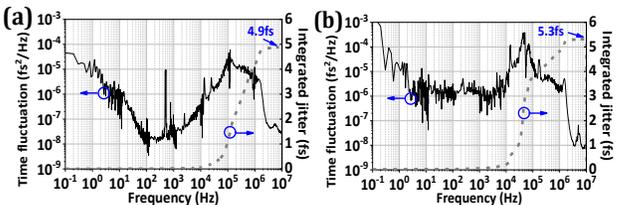

**Fig. 5.** Calculated time fluctuation PSDs (black solid curves) and the corresponding integrated timing jitters (grey dash curves) of (a) comb 1 and (b) comb 2.

**Table. 1.** Summary of performance of the two combs

| | Fast actuator | Resonance frequency | Implementation mode | Cross talk | $f_{ceo}$ phase error (0.1 Hz–10 MHz) | $f_{beat}$ phase error (0.1 Hz–10 MHz) | Frequency instability @ 1s | Timing jitter (0.1 Hz–10 MHz) |
|---|---|---|---|---|---|---|---|---|
| Comb 1 | EOM | 900 kHz | Transmission | Obvious | 0.49 rad | 0.35 rad | $1.5 \times 10^{-18}$ | 4.9 fs |
| Comb 2 | PZT | 180 kHz | Reflection | Negligible | 0.44 rad | 0.65 rad | $1.5 \times 10^{-18}$ | 5.3 fs |

The characteristics of the two fast actuators investigated in this work are summarized in Table. 1. The experimental results show that both the EOM and PZT are able to achieve the tight phase locking in high-repetition-rate fiber combs with the same level of frequency stability. An EOM suppresses noise with a larger servo bandwidth, but it is inhibited by competition with $f_{ceo}$, which leads to the jitter at high frequencies. On the other hand, the PZT does not interfere with $f_{ceo}$, but rather exhibits a relatively smaller servo bandwidth limited by the mechanics. However, outstanding performances is expected from the PZT by further optimizing the resonance frequency using a special damping alloy [26].

It should be emphasized that good locking performance of $f_{ceo}$ plays an important role for achieving tight phase locking of $f_{beat}$. This is because stabilization of $f_{ceo}$ suppresses noise that arises from the repetition rate [25]. The resulting free-running linewidth of $f_{beat}$ was remarkably narrowed when the $f_{ceo}$ loop was enabled (see Fig. 2(b)). The consequent benefit is the reduced bandwidth requirements of the servo loop for both the EOM and PZT. Therefore, even a response bandwidth of 180 kHz in the PZT is sufficient to phase lock $f_{beat}$.

In conclusion, we demonstrated two tightly phase-locked 750-MHz Yb:fiber combs stabilized to an optical reference with low residual phase errors. $f_{ceo}$ with ~200 kHz free-running linewidth was phase locked to exhibit a ~75 dB×Hz coherent peak using simple LD current modulation. Furthermore, an EOM and PZT were exploited for tight phase locking of $f_{beat}$ in such high-repetition-rate Yb:fiber combs. Both are able to suppress low frequency noise and reduce jitter to a few femtoseconds. The in-loop fractional frequency instability with either modulator can reach ~$1.5 \times 10^{-18}$ at 1 s. The experiment results prove that both are practical schemes for low noise operation of the frequency combs. The aforementioned techniques are powerful tools for achieving low noise with high repetition rate in Yb:fiber combs for advanced applications. In addition, one of the combs is also amplified to 10 W with sub-100 fs de-chirped pulse duration, and the noise properties will be investigated and reported elsewhere.

**Funding.** Japan Science and Technology Agency (JST) through the ERATO MINOSHIMA Intelligent Optical Synthesizer (IOS) Project (JPMJER1304); National Natural Science Foundation of China (61575004, 61735001).

**Acknowledgment.** The authors are in debt to Hajime Inaba of the National Institute of Advanced Industrial Science and Technology (AIST), Kenichi Nakagawa of the University of Electro-Communications (UEC), and Haochen Tian of Tianjin University (TJU) for their generous suggestions.

†These authors contributed equally to this work.